\documentclass[%
reprint,
superscriptaddress,
frontmatterverbose, 
 amsmath,amssymb,
aps,
]{revtex4-2}
\usepackage{xcolor}
\usepackage{graphicx}
\usepackage{dcolumn}
\usepackage{bm}
\usepackage{blindtext}
\usepackage{hyperref}

\usepackage[
scale=0.85, marginratio={1:1, 2:3}, ignoreall,
margin=0.65in,
]{geometry}

\begin{document}

\preprint{APS/123-QED}

\title{Universal Hydrogen Bond Symmetrisation Dynamics Under Extreme Conditions}

  \author{Thomas Meier}
  \email{thomas.meier@hpstar.ac.cn}
  \affiliation{Center for High Pressure Science and Technology Advance Research, Beijing, China}
   
  \author{Florian Trybel}
  \affiliation{ Department of Physics, Chemistry and Biology (IFM), Link{\"o}ping University, SE-581 83, Link{\"o}ping, Sweden} 
  
  \author{Saiana Khandarkhaeva}
  \affiliation{Bayerisches Geoinstitut, University of Bayreuth,  Bayreuth, Germany}
  
  \author{Dominique Laniel}
  \affiliation{Material Physics and Technology at Extreme Conditions, Laboratory of Crystallography, University of Bayreuth,  Bayreuth, Germany}

  \author{Takayuki Ishii}
  \affiliation{Center for High Pressure Science and Technology Advance Research, Beijing, China}

  \author{Alena Aslandukova}
  \affiliation{Bayerisches Geoinstitut, University of Bayreuth, Bayreuth, Germany}
  
  \author{Natalia Dubrovinskaia}
  \affiliation{Material Physics and Technology at Extreme Conditions, Laboratory of Crystallography, University of Bayreuth,  Bayreuth, Germany}                         
\affiliation{ Department of Physics, Chemistry and Biology (IFM), Link{\"o}ping University, SE-581 83, Link{\"o}ping, Sweden} 
  \author{Leonid Dubrovinsky}
  \affiliation{Bayerisches Geoinstitut, University of Bayreuth, Bayreuth, Germany}  

\date{\today}

\begin{abstract}
The experimental study of hydrogen bonds and their symmetrisation under extreme conditions is predominantly driven by diffraction methods, despite challenges of localising or probing the hydrogen subsystems directly. Until recently, H-bond symmetrisation has been addressed in terms of either nuclear quantum effects, spin crossovers or direct structural transitions; often leading to contradictory interpretations when combined. Here, we present high-resolution \textit{in-situ} $^1$H-NMR experiments in diamond anvil cells investigating a wide range of hydrogen bonded systems at pressure ranges of up to 90 GPa covering their respective H-bond symmetrisation. We found pronounced minima in the pressure dependence of the NMR resonance line-widths associated with a maximum in hydrogen mobility, precursor to a localisation of hydrogen atoms. These minima, independent of of the chemical environment of the linear O---H-O unit, can be found in a narrow range of oxygen-oxygen distances between 2.44 and 2.45 \AA, leading to an average critical oxygen-oxygen distance of  $\bar{r}_{\rm OO}^{crit}=2.443(1)$ \AA.
\end{abstract}

\maketitle

Understanding the stability and properties of hydrous minerals, possibly contributing to hydrogen transport to the lower mantle is crucial as key properties of the constituents of Earth's mantle, e.g., melting temperatures, rheology, electrical conductivity and atomic diffusivity \cite{bercovici2003,inoue1994,karato1986,yoshino2006} can be strongly affected by the presence of even small amounts of hydrogen. In particular, the high-pressure ($P$) phases of H$_{\rm 2}$O ice,  (Al,Fe)OOH and dense hydrous magnesium silicates are important candidates for hosting large amounts of hydrogen \cite{thompson1992}. These phases have a common O-H-O sequence, with an asymmetric hydrogen bond (O$\cdots$H-O) at low compression, which symmetrizes under increasing $P$ (O-H-O). During the symmetrisation, compressional (e.g. bulk modulus) as well as transport properties undergo significant modifications, but although the common  O$\cdots$H-O sequence endorses the conclusion that the phases should show very similar behaviour, significantly distinct characteristics are observed, e.g. (i) the  symmetrization $P$ has a great variation, e.g. $\simeq 120$ GPa in ice-VII \cite{Trybel2020,Mendez2021,benoit2002reassigning} compared to $\simeq 15$ GPa in $\delta$-AlOOH \cite{Trybel2021,pillai2018,cortona2017}, (ii) proton tunnelling dominates the transition in ice-VII \cite{Trybel2020, Meier2018,Mendez2021, lin2011correlated, Goncharov1996}, but is absent in $\delta$-AlOOH \cite{Trybel2021}. (iii) The bulk modulus in ice-VII \cite{Mendez2021,li2019high,ahart2011brillouin} shows a noticeable softening, while only a minor effect is found in $\delta$-AlOOH \cite{pillai2018} .
\begin{figure}[!h]
\begin{center}
\includegraphics[width=0.45\textwidth]{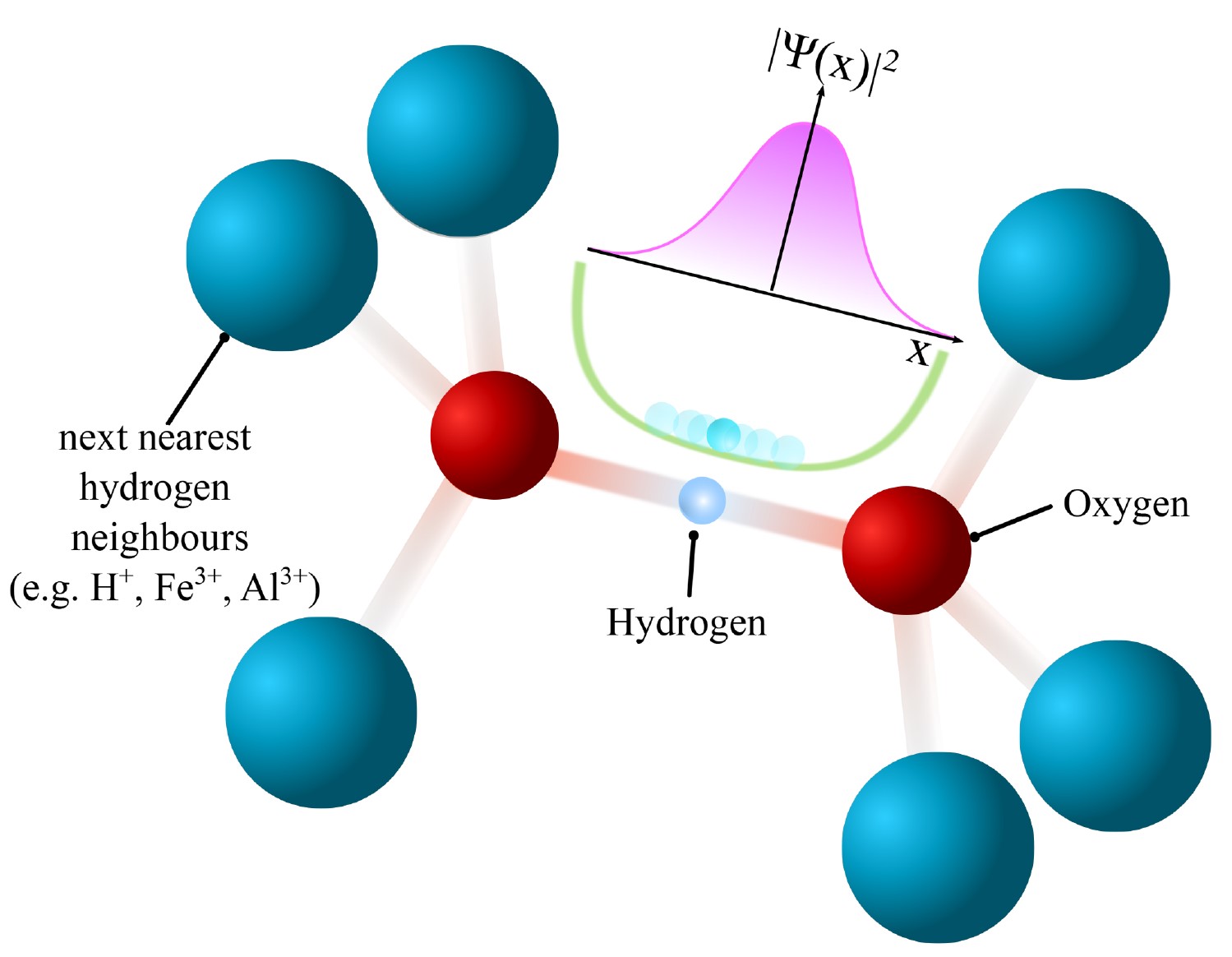}%
\caption{\textbf{Structure of hydrogen bond environments.} \textbf{a)} Schematic representation of the local hydrogen bond environments with the hydrogen atom at the center. The energy potential of the H-bond is determined predominantly by the nearest neighbour atoms (e.g. oxygen atoms). Depending on initial O-O distances, hydrogen probability distributions (purple schematic graph) may exhibit uni- or bi-modal characteristics. Shown here is a typical broad energy potential without barrier, leading to an uni-modal probability distribution and thermally activated hydrogen mobility. Next nearest neighbours may constitute metal anions, e.g., in hydrous minerals, or other hydrogen atoms in H$_{\rm 2}$O ice-VII/X.}
\label{fig:structure}
\end{center}
\end{figure} 

In order to shed light on the underlying mechanisms it is essential to gain a deeper understanding of the similarities and disparities between different oxide-hydroxide phases with respect to the hydrogen-bond symmetrization. Hydrogen atoms, however, have a very low X-ray cross-section and neutron techniques are typically unavailable at $P \gtrsim 25$ GPa, therefore knowledge about the hydrogen subsystem is widely limited to computational and optical studies (e.g. Raman or Infrared). Advances in high-$P$ nuclear magnetic resonance spectroscopy in recent years \cite{Meier2015a,Meier2018b,Meier2021}, allow for a direct investigation of the hydrogen subsystem (Figure \ref{fig:structure}) beyond the megabar-range and enables to obtain previously unattainable experimental insight.
\onecolumngrid

\begin{figure*}[h!]
\includegraphics[width=0.85\textwidth]{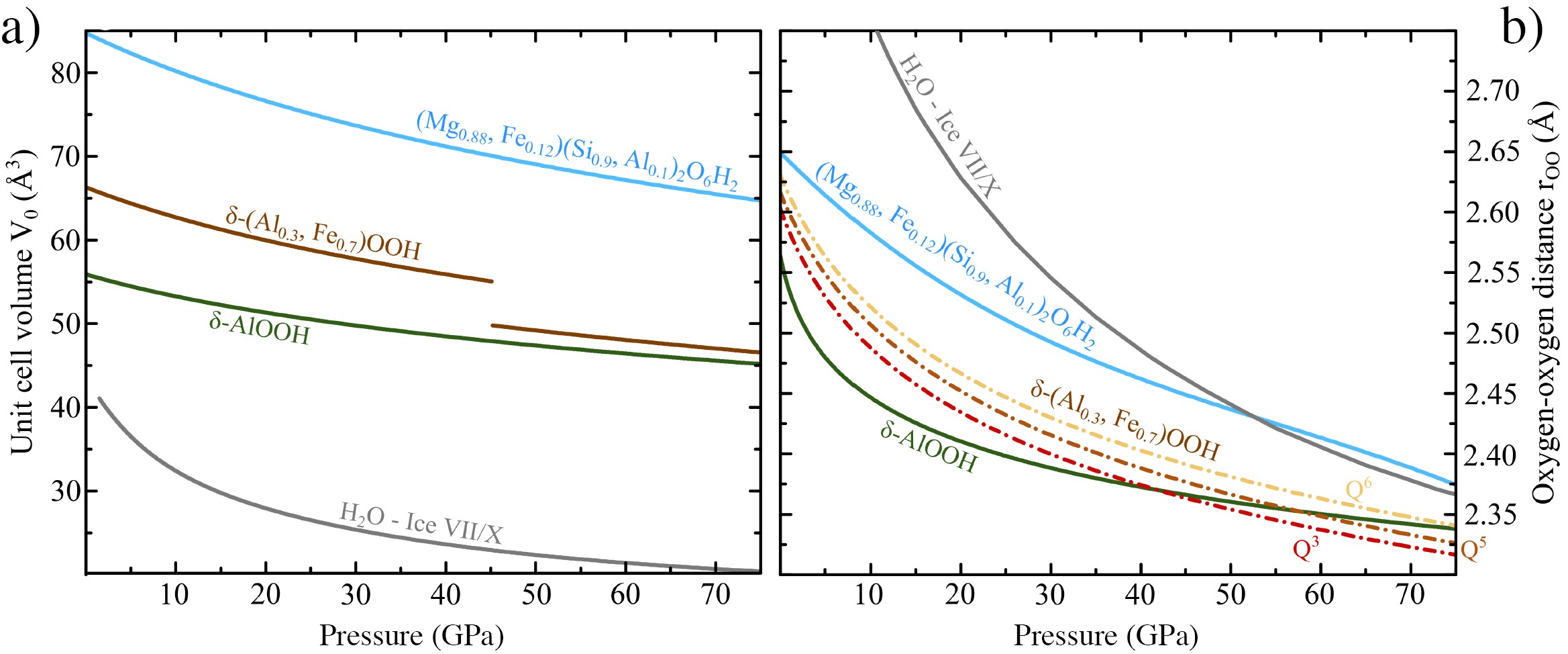}%
\caption{ \textbf{a)} Equation of state (EOS) data of the different hydrogen bonded phases. We use the EOS for H$_2$O ice-VII/X of from \citet{French2015}, based on \textit{ab-initio} computations and the XRD based EOS of \citet{Simonova2020} for $\delta$-AlOOH. The other curves are based on third order Birch-Murnagham EOS fits to our diffraction data for iron-bearing $\delta$-AlOOH and phase-D. $\delta$-(Al$_{0.3}$,Fe$_{0.7}$)OOH exhibits a 7\% volume collapse associated to the electron spin crossover of ferric iron (\textit{c.f.} Fig. \ref{fig:FWHMshift}a). \textbf{b)} Oxygen-oxygen distance ($r_{\rm OO}$) as a function of $P$. Different pressure dependencies of the $Q^i$(i=3,5,6) H-bond environments in $\delta$-(Al,Fe)OOH \cite{Meier2021} were estimated through interpolation of $r_{\rm OO}$ between endmembers $\delta$-AlOOH \cite{Trybel2021,Simonova2020} and $\epsilon$-FeOOH \cite{Xu2013} with respect to the local iron content. We only consider the high spin state in iron bearing phases, as the spin transition occurs at significantly higher $P$ ($\sim 45$ GPa) than the hydrogen-bond symmetrization ($\sim 15$ GPa)}
\label{fig:EOS}
\end{figure*}
\twocolumngrid

Here, we present \textit{in-situ} high-$P$ $^{1}$H-NMR data on four hydrous mineral phases: (i) dense magnesium silicate phase-D with composition (Mg$_{0.88}$,Fe$_{0.12}$)(Si$_{0.9}$, Al$_{0.1}$)$_2$O$_6$H$_2$, (ii) pure and (iii) iron-bearing aluminum oxide-hydroxide ($\delta$-(Al$_{\rm0.3}$,Fe$_{\rm 0.7}$)OOH) as well as (iv) high-$P$ phases of H$_2$O (ice-VII/X). We find a distinct maximum in hydrogen mobility in all sampled systems at the same critical O$\cdots$H-O in-bond oxygen-oxygen distance ($r_{\rm OO}^{\rm crit}$). Being a precursor for H-bond symmetrisation and under further compression localisation, this effect was found to solely depend on the short-range local structure, \textit{i.e.} $r_{\rm OO}$ is independent of the nature of the further chemical surrounding, marking $r_{\rm OO}^{\rm crit}$ as an unifying parameter of pressure-induced hydrogen bond symmetrisation.

We use the  equation of state (EOS) parameters (Figure \ref{fig:EOS}a) for $\delta$-AlOOH ($V_{\rm 0}=56$ \AA$^{3}$, $K_0= 183$ GPa,$K_0'= 3.7$)
from \citet{Simonova2020},  $\delta$-(Al$_{0.3}$,Fe$_{0.7}$)OOH ( $V_{\rm 0}=63.71$ \AA$^{3}$,  $K_0= 164.7 $ GPa, $K_0'= 4.04$)  and Phase-D ($V_{\rm 0}=84.73$ \AA$^{3}$, $K_0= 162$ GPa, $K_0'=4$), are determined from synchrotron X-ray diffraction (XRD, see Methods section). For  H$_2$O ice-VII/X we use the global EOS by \citet{French2015}, based on DFT calculations, which was found to be in excellent agreement with recent dynamical diamond anvil cell XRD experiments \cite{Mendez2021}. The volume collapse in iron bearing oxide-hydroxide at $P\approx45$ GPa originates from the $S=5/2$ high spin to $S=1/2$ low spin transition of ferric iron \cite{hsieh2020spin,thompson2020phase}. Analogous transitions in dense magnesium silicate phase-D are not observable in our XRD data due to the low concentration of iron, but distinct transitions can be found in synchrotron M\"ossbauer spectroscopy (SMS) (see Methods section).

In order to quantify the evolution of hydrogen NMR signals with respect to the \textit{local} symmetry under compression, we calculate oxygen-oxygen distances in the H-bonds ( $r_{\rm OO}$), using both diffraction data (for $\delta$-(Al,Fe)OOH \cite{Simonova2020} and phase-D) as well as DFT calculations for high-$P$ ice-VII/X \cite{French2015}.  As shown recently in ref. \cite{Meier2021}, the hydrogen bond manifold in $\delta$-(Al$_{0.3}$,Fe$_{0.7}$)OOH can be resolved and the observed signals assigned to different local environments, $Q^{i}$~$(i=1-6)$, where the fraction $i$ of 6 possible next nearest neighbour positions around the hydrogen bond (Fig \ref{fig:structure}) is occupied by ferric iron paramagnetic centres. We deduce the $P$ dependence of $r_{\rm OO}$ in $\delta$-(Al$_{0.3}$,Fe$_{0.7}$)OOH by interpolating between the endmember $\delta$-AlOOH \cite{Trybel2021,Simonova2020} and $\epsilon$-FeOOH \cite{Xu2013} with respect to the local iron content around the hydrogen nearest neighbours (Fig. \ref{fig:structure}). We neglect the volume collapse due to the spin transition in iron bearing phases as is at significantly higher $P$ than the hydrogen-bond symmetrization.

\begin{figure*} 
\includegraphics[width=0.8\textwidth]{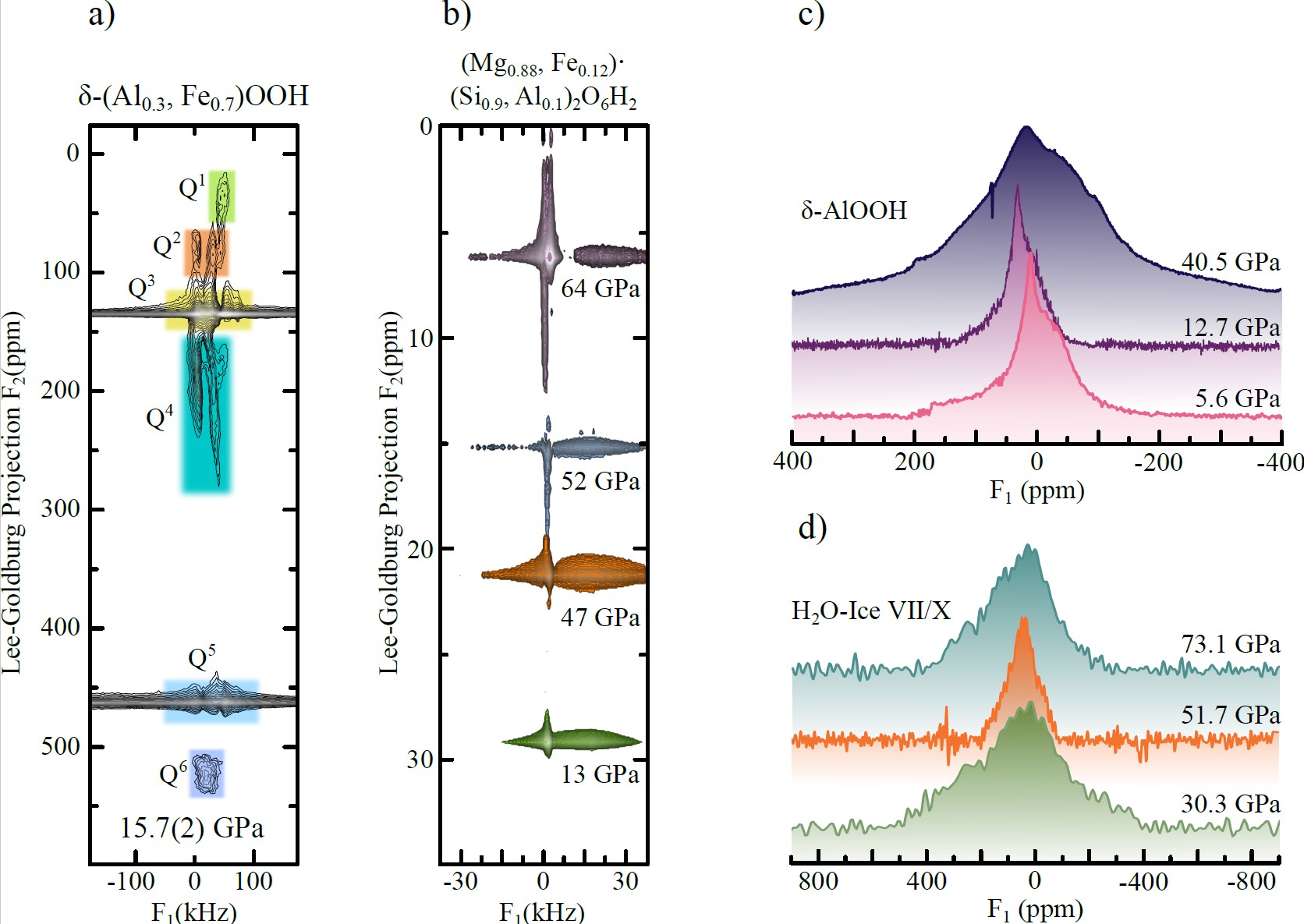}%
\caption{\textbf{$^1$H-NMR spectra of all sample systems:} \textbf{(a)} High-resolution 2D-Lee Goldburg (LG) spectrum of $\delta$-(Al$_{0.3}$,Fe$_{0.7}$)OOH at 15.7 GPa showing up to six distinguishable hydrogen signals in the indirect LG projection $F_2$ \cite{Meier2021}. Assuming a stochastic distribution of ferric iron in the sample, a comparison with signal intensities allows for the assignment of the H-bond species $Q^i$(i=1-6). Spectra at different pressures were recorded at reduced resolution due to experimental time restrictions and solely signals from $Q^3,~Q^5$ and $Q^6$ species were recorded over the full pressure range. \textbf{(b)} High-resolution 2D-LG spectra of dense magnesium silicate phase-D, showing a shift towards lower chemical shifts at the HS$\rightarrow$LS crossover (fig. \ref{fig:FWHMshift}a). Measurements were conducted at a magnetic field of 7.04 T.  $^1$H-NMR solid-echos of \textbf{(c)} $\delta$-AlOOH and  \textbf{(d)} H$_2$O ice-VII/X as a function of pressure. Measurements were conducted at a magnetic field of 1.2 T.}
\label{fig:spectra}
\end{figure*}

We performed high-resolution $^1$H-Lee Goldburg (LG) NMR measurements \cite{Lee1965a} for $\delta$-(Al$_{0.3}$,Fe$_{0.7}$)OOH and Mg$_{0.88}$,Fe$_{0.12}$)(Si$_{0.9}$, Al$_{0.1}$)$_2$O$_6$H$_2$ phase-D in order to investigate the influence of different next nearest neighbour configurations (Figure \ref{fig:structure}) and recorded $^{1}$H-NMR solid echos for pure $\delta$-AlOOH as well as H$_2$O ice-VII/X. 

The electronic environment of the hydrogen nuclei should, to first approximation, be determined by their nearest neighbours, \textit{i.e.} the oxygen atoms involved in the hydrogen bonds \cite{Slichter1978}. Presence of paramagnetic centres among the next nearest neighbours, such as in phase-D or $\delta$-(Al$_{0.3}$,Fe$_{0.7}$)OOH, induces additional contributions to NMR resonance shifts and relaxation times \cite{Pell2019}. These should not significantly influence dynamical effects, but instead lead to a modulation of the energy landscape of H-bonds in these systems \cite{Levitt2000} and therefore a separation of the signals in the indirect LG-projection frequency domain $F_2$ is observable.

Figure \ref{fig:spectra}a shows a high-resolution $^1$H-LG NMR spectrum of $\delta$-(Al$_{0.3}$,Fe$_{0.7}$)OOH at 15.7 GPa. As recently shown \cite{Meier2021}, the 2D LD spectrum can be separated in six different regions. Under the assumption of a stochastic distribution of ferric iron, several different local H-bond environments should be present in the sample due to a second order modification of the electronic environment. We associated these different region to the different next-nearest neighbour surroundings ($Q^i$), by matching the relative size of the regions with the probability distribution of finding a local surrounding with $i$ of 6 possible positions being occupied by iron atoms \cite{Meier2021}. Signals associated to fully Fe$^{3+}$ depleted H-bond environments ($Q^0$) lie below the detection limit of our experiments. Interestingly, the signal associated to four ferric iron paramagnetic centres occupying next-nearest neighbour positions ($Q^4$) was found to be much broader than to the other H-bond environments, possibly due to significant gradients in the local magnetic field stemming from an asymmetric distributions of paramagnetic centres, further hyperfine interactions or signal overlap \cite{Pell2019}. The spectrum shown in Figure \ref{fig:spectra}a was recorded by oversampling in the indirect Lee-Goldburg dimension $F_2$, other spectra were recorded using faster $F_2$ sampling and only the most intense signals associated to $Q^3$, $Q^5$ and $Q^6$ were recorded (see Supplementary Material).

\begin{figure*}
\begin{center}
\includegraphics[width=0.85\textwidth]{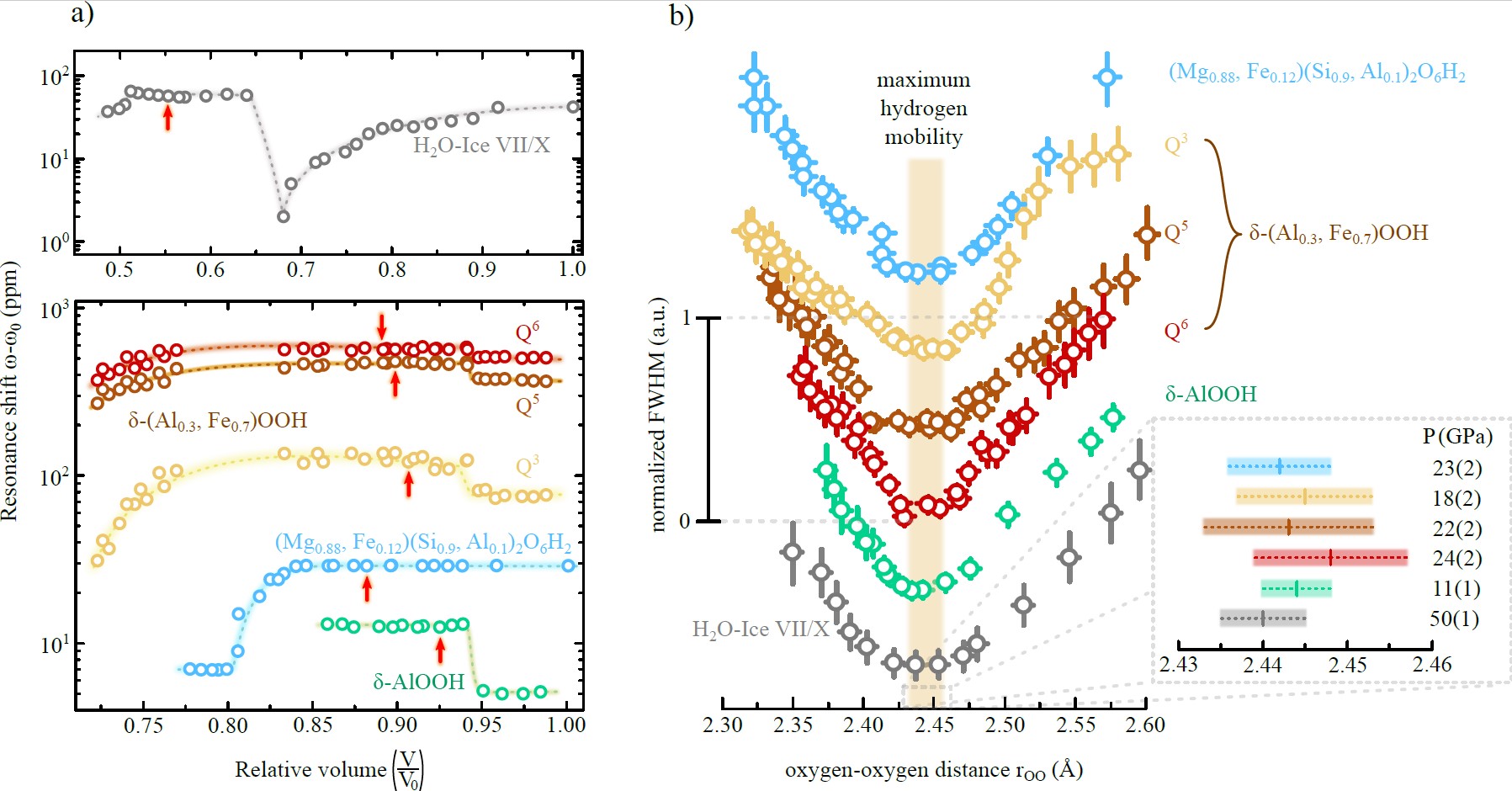}%
\caption{\textbf{Resonance shift and FWHM line widths.}
\textbf{a) Top panel:} Resonance shift found in $^1$H-NMR solid echos of H$_{\rm 2}$O ice VII/X. The signal shift of the  resonances follows the dependence previously reported at high magnetic fields \cite{Meier2018}: after an initial decrease up to 20 GPa ($V/V_0\approx0.68$), the signals are shifted downfield corresponding to the transition from a high to low barrier hydrogen bond regime. The second jump at 73 GPa ($V/V_0\approx0.52$) signifies the onset of the ice-VII $\rightarrow$~X crossover.
\textbf{Lower panel:} Resonance shift of $\delta$-AlOOH, $\delta$-(Al,Fe)OOH and phase-D as a function of the relative volume. Both $\delta$-AlOOH and $\delta-$(Al$_{0.3}$,Fe$_{0.7}$)OOH undergo a sub to supergroup phase transition from $P2_{1}nm$ to $Pnnm$ which was observed at $V/V_0\approx0.94$ by a sudden increase of the shift by about 5 - 10 ppm. Between  $V/V_0 \approx 0.83 - 0.81$, both iron bearing $\delta$-(Al$_{0.3}$,Fe$_{0.7}$)OOH as well as phase-D undergo an electron spin crossover, resulting in a partial reduction of the paramagnetic shift interaction as well as a volume collapse observed in $Q^3$, $Q^5$ and $Q^6$. The red arrows mark the points of maximum hydrogen mobility. \textbf{b)} Normalized FWHM line widths as a function of $r_{\rm OO}$. All investigated systems go through a maximum of proton mobility, \textit{i.e.} motional narrowing of resonance lines and a consequent minimum of FWHM line widths at critical oxygen-oxygen distances between $r^{\rm crit}_{\rm OO}=2.44$ and $2.45$, with an average value of  $\bar{r}^{\rm crit}_{\rm OO}=2.443(1)$ \AA.   
}
\label{fig:FWHMshift}
\end{center}
\end{figure*}

At ambient conditions, the resonance frequencies of these signals were found to be $\omega-\omega_0=~80,~365$ and $506$ ppm for $Q^3$, $Q^5$ and $Q^6$, respectively (Fig. \ref{fig:FWHMshift}a). The origin of such a pronounced proton resonance shift can reasonably be expected to be hyperfine interaction of the hydrogen nuclei with localised electron moments of Fe$^{3+}$ ions, which should be modulated at the electron spin crossover around $V/V_0\approx 0.85$ to $0.80$. Indeed, we find a significant reduction in resonance frequency of all three signals at $V/V_0= 0.833$, in excellent agreement with M\"ossbauer spectroscopy (see Methods section), evidencing the paramagnetic origin of these signals. Additionally, we observe a discontinuity in resonance shift at a relative compression of $V/V_0= 0.94$ (Fig. \ref{fig:FWHMshift}a) in all $Q^{\rm i}$ and therefore independent of the local iron content, which we associate with the sub- to super-group phase transition from $P2_{1}nm$ to $Pnnm$ at $P\simeq 10$ GPa \cite{Simonova2020, Trybel2021,Sano-Furukawa2018}.

${^1}$H-NMR signals of dense magnesium silicate phase-D (Fig. \ref{fig:spectra}b) show a narrow resonance in the indirect LG-projection frequency domain $F_2$, indicating a single, well defined coordination of hydrogen atoms. The observed frequency shift of the signals (Fig. \ref{fig:FWHMshift}a) towards lower ppm values coincides with the electron spin crossover of ferric iron Fe$^{3+}$ in $\delta$-(Al$_{0.3}$,Fe$_{0.7}$)OOH. Therefore, we assume the observed signals to stem from hydrogen atoms predominantly influenced by the presence of Fe$^{3+}$ ferric iron in phase-D. Signals shifted by the presence of Fe$^{2+}$ ferrous iron (see Methods section) were not observed in the chosen spectral range.

One-dimensional solid echo NMR spectra of pure $\delta$-AlOOH (Fig. \ref{fig:spectra}c) at 1.2 T show a single signal at $\omega-\omega_0\approx$ 0 ppm close to the anticipated position of ferric iron depleted $Q^i$ species in the iron bearing oxide-hydroxide analogue, indicating a well defined hydrogen position in agreement with earlier NMR studies under ambient conditions \cite{Xue2007}. Similar to $\delta$-(Al$_{0.3}$,Fe$_{0.7}$)OOH, we associate the discontinuity at $V/V_0=0.94$ (Fig. \ref{fig:FWHMshift}a) to the aforementioned sub- to super-group phase transition. 

Figure \ref{fig:spectra}d shows representative $^1$H-NMR spectra of H$_2$O ice-VII/X recorded at 1.2 T. We observe two distinct transitions in the $P$ dependence of the resonance frequencies (Fig. \ref{fig:FWHMshift}a): (i) At $V/V_0=0.66$ (20 GPa), after a continuous shift of about 40 ppm up-field, the proton signals showed a sudden discontinuity of $\approx 55$ ppm, indicating the transition from the high to low barrier hydrogen bond regime in ice-VII, where the number of tunnelling protons is increasing \cite{Trybel2020,Mendez2021} and (ii) at $V/V_0=0.51$ (73 GPa) a discontinuity of $\approx$ 20 ppm associated with the beginning of the continuous transition from ice-VII to ice-X. This behaviour is in excellent agreement with our previous study \cite{Meier2018}.

In order to find a common parameter indicating the hydrogen bond symmetrization, we calculate the full width at half maximum (FWHM) line-widths as a function of $P$ as well as relative compression ($V/V_0$). We find a minimum in the $^1$H-NMR FWHM line-widths in all signals. These minima occur within a wide compression range, \textit{i.e.} $V/V_0\approx~0.92$ to $0.55$ (red arrows in Fig. \ref{fig:FWHMshift}a), corresponding to $P$ between 13 to 50 GPa. We furthermore calculate the FWHM as a function of the respective oxygen-oxygen distances ($r_{\rm OO}$) for all four compounds, resolving different local environments ($Q^{i}$) in $\delta-$(Al$_{0.3}$,Fe$_{0.7}$)OOH (Figure \ref{fig:FWHMshift}b). Surprisingly, all signals were found to have a continuous transition through this pronounced minimum at almost identical O-O distances $r^{\rm crit}_{\rm OO}=2.44$ and $2.45$, with an average value of  $\bar{r}^{\rm crit}_{\rm OO}=2.443(1)$ \AA. Several possible error sources might lead to a minor modulation of this value, such as insufficient diffraction data above 40 GPa in the electronic low spin state or the interpolation of the $r_{\rm OO}$ evolution of the $Q^i$ species in $\delta$-(Al$_{0.3}$,Fe$_{0.7}$)OOH. Nevertheless, estimating errors based on these effects was found to not significantly alter the position of the observed minima but only the respective width and thus the spread of $r^{\rm crit}_{\rm OO}$ (\textit{c.f.} inset in Fig. \ref{fig:FWHMshift}b). 
In absence of magic angle spinning \cite{Hennel2005} or Lee-Goldburg derived experiments \cite{Meier2021}, line narrowing in NMR mainly originates from two distinct mechanisms: (i) local structural symmetrisation, leading to sharper resonance line distributions (structural narrowing) and (ii) local diffusive motion of the probed nuclei leading to an averaging of short range electromagnetic interactions (dynamic narrowing) \cite{Slichter1978}.

While the first effect is particularly pronounced in quadrupolar NMR as the electric nuclear quadrupole moment of any $I>1/2$ nucleus is a sensitive probe of the local electronic environment and charge distribution; such an effect might not be very pronounced for $I=1/2$  (\textit{e.g.} $^1$H) nuclei. Contrarily, moderate dynamic averaging of hydrogen NMR resonances is a well established diagnostic tool to identify locally restricted low amplitude motions  and has even been employed to determine hydrogen diffusivities in diamond anvil cell based research \cite{Meier2020a}. This minimum indicates the point of maximum hydrogen diffusivity within the H-bond, leading to a striking conclusion. At low pressures and long $r_{\rm OO}\approx 2.60-2.50$ \AA, hydrogen mobility continuously increases. At $r^{\rm crit}_{\rm OO}$, hydrogen mobility reaches a maximum, \textit{i.e.} minimum in the FWHM line width, as the majority of protons de-localise.

Even more interesting is the fact that the point of maximum hydrogen de-localisation in the oxide-hydroxides and phase-D, possessing a single-well H-bond energy potential is nearly identical to the maximum found in H$_2$O ice-VII/X characterised by a pronounced double-well character and correlated proton tunnelling \cite{Lin2011,Trybel2020} . This indicates that the observed effect is in a region of $r^{\rm crit}_{\rm OO}$ where the barrier of the double-well potential is already negligible compared to the protons energy and therefore that the high-$P$ regime is independent of the low-$P$ symmetry of the H-bond  potential and associated tunnelling. 

Being a primer for hydrogen bond symmetrisation or, more accurately, for hydrogen localisation as the diffusive motion of protons diminishes to $r_{\rm OO}<r^{\rm crit}_{\rm OO}$, this hydrogen mobility maximum is spread over a surprisingly wide range of over 40 GPa, going from 13 GPa in $\delta$-AlOOH to about 50 GPa in H$_2$O ice-VII/X. Since the FWHM linewidth dependencies are comparable in all observed samples, despite significantly different charge distributions of each environment (ration of Al and Fe in $Q^{i}$, O, etc.) as well as 3D (e.g. H$_2$O ice-VII/X) and 2D (e.g. $\delta$-AlOOH) hydrogen bond networks, it is reasonable to assume that this effect is universal for all linear O-H-O hydrogen bonded materials regardless of their long range atomic structure.  

Furthermore, it has been argued that H-bond symmetrisation or hydrogen localisation might either be a primer to electron spin crossovers in iron bearing hydrous minerals like oxide-hydroxides\cite{Xu2013, Gleason2013a}. Contrary, our NMR data clearly demonstrates that electron spin transitions and pressure induced hydrogen localisation are independent physical phenomena.

In this study, we have shown that hydrogen bond symmetrisation and proton localisation as a consequence of the modulation of the H-bond energy potential is a phenomenon inherent to wide variety of materials ranging from high-$P$ H$_2$O ices to (iron-bearing) oxide-hydroxides. The observed hydrogen bond symmetrisation and proton localisation dynamics were found to follow identical scaling behaviour in the O-H-O oxygen-oxygen distance $r_{\rm OO}$ and are solely dependent on the short range atomic and electronic structure. We identified an average critical oxygen-oxygen distance of $2.443(1)$ \AA~ where the delocalisation of hydrogen atoms in H-bonds is maximised. 
\section*{Methods}
\subsection{Sample preparation and characterisation}

\subsubsection{{\rm H$_2$O ice-VII/X \& $\delta$-AlOOH}}
The sample preparation and measurement of the H$_2$O ice-VII/X and $\delta$-AlOOH data is in detail described in references \cite{Meier2018a}  and \cite{Trybel2021,Simonova2020}.

\subsubsection{{\rm $\delta$-(Al$_{0.3}$,Fe$_{0.7}$)OOH}}

A mixture of regeant grade FeOOH and Al(OH)$_3$ with a molar ratio of $7:3$ was used as starting material. Single crystals of $\epsilon$-(Fe,Al)OOH were synthesized using a 1000-ton Kawai-type multi-anvil high-pressure apparatus at Bayerisches Geoinstitut, University of Bayreuth, Germany. Tungsten carbide (WC) anvils with 4-mm truncated edge lengths were used to compress the sample in combination with a $5$ wt\%  Cr$_2$O$_3$-doped MgO octahedral pressure medium with a 10 mm edge length. A cylindrical LaCrO$_3$ heater was set at the center of the pressure medium. The starting material was packed in a platinum capsule, which was welded shut. The sample capsule was inserted in an MgO capsule and put at the central part of the heater. Sample temperature was monitored at the central part of the outer surface of the $Pt$ capsule using a W$-3\%$Re/W$-25\%$Re thermocouple, neglecting pressure effects on electromotive force of the thermocouple. \\
The sample was compressed to a desired press load at room temperature and then heated to a target temperature of 1200$^{\circ}$C at a rate of 100$^{\circ} C/$min. After keeping this temperature for 180 min, the sample was quenched by turning off electrical power and slowly decompressing to ambient pressure for 15 hours. 
\\
Recovered single crystals with dimensions up to 200 $\mu$m were selected based on the absence of twinning and sharp optical extinction using a polarizing microscope. The crystals were mounted on a glass fibre and  X-ray diffraction (XRD) data collected using a Huber Eulerian cradle single-crystal diffractometer driven by the SINGLE software \cite{Angel2011}. The diffractometer was equipped  with a Mo K$\alpha$ X-ray source and  operated at 50 kV and 40 mA. Crystals with a half-width of the diffraction peaks less than 0.1$^{\circ}$ were further analysed in terms of their cell parameters with the vector least-squares method \cite{Ralph1982}. The effect of crystal offsets and diffractometer aberrations for each crystal was eliminated using the eight-position centring method \cite{King1979}. The crystal had the space group of $P2_1nm$ and lattice parameters of a = 2.9573(1) \AA, b = 4.3884(1) \AA, c = 4.8873(2), V = 63.426(4) \AA$^3$. 
\\
Chemical compositions of the samples were measured using an electron microprobe analyzer (EPMA) with wavelength-dispersive spectrometers (JEOL, JXA-8200) operated at 15 kV and 10 nA, for 20 s on the peaks of Al and Fe and 10 sec on the background. Synthetic hematite and corundum were used as standard material for Fe and Al, respectively. The composition of the dark-red crystals was expressed as (Al$_{0.305(9)}$,Fe$_{0.695(9)}$)OOH, assuming that the cation number of hydrogen is one.
\\

\subsubsection{{\rm(Mg$_{0.88}$,Fe$_{0.12}$)(Si$_{0.9}$,Al$_{0.1}$)$_2$O$_6$H$_2$ phase-D}}

A starting material was prepared as a powdered mixture of regent grade chemicals of Mg(OH)2 (45.11 wt.\%), SiO2 (38.02 wt\%), Al(OH)3 (5.48 wt.\%) and $^57$Fe enriched Fe$_2$O$_3$ (11.38 wt.\%). Single crystals of hydrous phase-D were synthesized using a 1200-ton Kawai-type multi-anvil high-pressure apparatus at Bayerisches Geoinstitut, University of Bayreuth, Germany . Tungsten carbide anvils with 3 mm truncated edge lengths were used in combination with a 5 wt\% Cr$_2$O$_3$-doped MgO octahedral pressure medium with a 7-mm edge length. A cylindrical LaCrO$_3$ heater was located at the center of the pressure medium. The starting material was packed in a welded platinum capsule. The sample capsule was inserted in an MgO capsule and put at the central part of the heater. Sample temperature was monitored at the central part of the outer surface of the Pt capsule using a W-3\%Re/W-25\%Re thermocouple. Pressure effect on electromotive force of the thermocouple was ignored. 

The sample was compressed to a desired press load at room temperature and then heated to a target temperature of 1100 ºC at a rate of 100 ºC/min. After keeping this temperature for 240 min, the sample was quenched by turning electrical power off and slowly decompressed to ambient pressure for 15 hours. 
Chemical composition of the recovered hydrous phase-D was determined using an electron microprobe analyzer (EPMA) with wavelength-dispersive spectrometers (JEOL, JXA-8200) operated at 15 kV and 10 nA, for 20 sec on the peaks of Al and Fe and 10 sec on the background. Natural enstatite for Mg and Si, synthetic hematite for Fe and corundum for Al were used as standard material.

\subsection{Mössbauer Spectroscopy}
Mössbauer absorption spectra were collected at ambient temperature in a diamond anvil cell at the Nuclear Resonance beamline (ID18) at the European Synchrotron Radiation Facility (Grenoble) using Synchrotron Moessbauer (SMS) Source spectroscopy\cite{Potapkin2012}. The experiment was conducted in transmission geometry and folded spectra contained 512 channels. The line width of the SMS was determined before and after collection of each pressure point by measuring the reference single line absorber (K$_2$Mg$_{57}$Fe(CN)$_6$). The Mössbauer spectra were fitted using MossA software\cite{Prescher2012} with the full transmission integral assuming a Lorentzian-squared line shape of the SMS. Isomer shift values are referred to that of $\alpha$-Fe at 300 K.
\\
$^{57}$Fe Mössbauer spectra of phase-D measured at ambient pressure and temperature show broadened asymmetric paramagnetic doublets, indicating a superposition of several subspectra. The spectra were interpreted as a mixture of two quadrupole doublets. The value of the isomer shift $\delta = 0.36(1) $mm/s of one subspectrum corresponds to high-spin (HS) ions Fe$^{3+}$ $(d5, S = 5/2)$ located in oxygen octahedra. Sub-spectra showing high values of isomer shifts $\delta = 1.22(2)$ mm/s are characteristics of  high-spin ferrous iron Fe$^{2+}$ $(d6, S = 2)$ \cite{Menil1985}. The high quadrupole splittings $\Delta$ of the doublets $0.81(5)$ mm/s and $2.19(3)$ mm/s showed that iron ions are located in crystal positions with a strong electric-field gradient, characteristic to a Al/Mg substitution in the phase-D crystal lattice.
\\
Upon compression, SMS spectra change drastically: above 25 GPa, spectra can no longer be described by only two doublets. Additional sub-spectra correspond to low-spin ferrous iron, indicating the onset of a HS $\rightarrow$ LS spin crossover. The ferric to ferrous iron ratio was constant throughout the whole compression range.
\\
Above 36 GPa, SMS signatures of HS ferric iron are lost and new sub-spectra with characteristic hyperfine parameters of LS Fe$^{3+}$ appear indicating another long range spin crossover. Isomer shifts LS Fe$^{3+}$ are lower than in the high-spin state and quadrupole splitting was found amplified due enhanced electric field gradients, associated with the non-uniform distribution of uncompensated $3d-t_{2g}$ valence electrons. The pressure dependence of hyperfine parameters show no anomalous changes in the local  environment of the iron ions.

\subsection{NMR-DAC preparation}

DACs for high-pressure NMR experiments were prepared following a procedure described in e.g  \cite{Meier2017,Meier2019a}. First, rhenium gaskets were indented to the desired thickness, which depends on the size of the diamond anvil culets employed, but usually $\lesssim25 \mu$m. Sample cavities were drilled using specialized laser drilling equipment. After gasket preparation, the diamond anvils were covered with a layer of 1 $\mu$m of copper or gold using chemical vapour deposition. To ensure the electrical insulation of the conductive layers from the rhenium gasket, the latter were coated by a thin layer ($\approx 500$ nm) of Al$_2$O$_3$ using physical vapour deposition. The Lenz lens resonators were shaped from the conductive layer on the diamonds by using focused ion beam milling. 
\\
Before the final cell assembly, radio frequency resonators were prepared accordingly to their desired operation frequency. A pair of high inductance solenoid coils ($\approx 100$ nH) for low frequency experiments of below 100 MHz or a pair of single turn printed circuit board (PCB) plated copper resonators for $^{1}$H-NMR frequencies at high fields were used as driving coil arrangements for the Lenz lens resonators' structure and were placed around each diamond anvil. After sample loading and initial pressurisation, the driving coils were connected to form a Helmholtz coil-like arrangement.
\\
Pressure calibration was performed using the shift of the first derivative of the first order Raman signal of the diamond edge in the center of the culet \cite{Akahama2004, Akahama2006}. All DACs were fixed and connected to home built NMR probes equipped with customized cylindrical trimmer capacitors (dynamic range of $\approx150$ pF) for frequency tuning to the desired resonance frequencies and impedance matching to the spectrometer electronics ($50~\Omega$). 
\\
Proton shift referencing was conducted using the $^{63}$Cu resonances of the Lenz lenses themselves as internal references taking into account the additional shielding of $B_0$ inherent to every DAC. These resonances were cross referenced with standard metallic copper samples at ambient conditions without a DAC. The resulting shift between both $^{63}$Cu-NMR signals are then used as a primer for the NMR signals of the samples under investigation.
\\
Lee-Goldburg decoupling experiments were initially calibrated by quick two-dimensional nutations ($\sim$ 56 2D spectra) for different off resonant frequencies of the LG-pulse. Optimal pulse offset frequencies $\omega_{\rm off}$ were found between 25 to 35 kHz. One dimensional LG-spectra were recorded by oversampling in the indirect time domain ($\sim$ 8000 increments) using previously determined values of $\omega_{\rm off}$ at a 10 dB pulse power attenuation relative to the excitation pulse. Two dimensional LG spectra were recorded with identical direct and indirect time domains (usually 2048 points in each dimension), while matching the incrementation of the LG-pulse to the direct time domain dwell time of the spectrometer. 

\subsection{X-ray diffraction}

Single crystal X-ray diffraction data for phase-D and $\delta$-(Al$_{0.3}$,Fe$_{0.7}$)OOH were collected at the beamlines P02.2 (PETRA III, DESY, Hamburg, Germany) and ID15 (ESRF, Grenoble, France). At PETRA III the data were collected using Perkin Elmer XRD1621 flat panel detector and X-rays with the wavelength $\lambda=0.2887$ \AA. For the different experiments at ID15 the beamline was equipped with a MAR555 flat panel or a EIGER2 X 9M CdTe (340x370 mm) detector. Before collecting XRD data, the experimental geometry was precisely defined based on freshly collected diffraction data from enstatite (at PETRA III) or a vanadinite callibrant (ESRF). The X-ray energy at ID15 was 30 keV and the wavelength is $\lambda$=0.4133 \AA, while the beamsize was $\sim$10 $\mu$m in FWHM. In all our experiments, at each pressure point data collection was performed in an omega range of least  $\pm$32$^{\circ}$ with 0.5$^{\circ}$ step per frame and an exposure time of 1 to 3 seconds per step. Integration of the reflection intensities and absorption corrections were performed using the CrysAlisPro software. Structure solution and refinement were done in the isotropic approximation using Jana2006\cite{petvrivcek2014crystallographic} with Superflip \cite{palatinus2007superflip} and SHELXT \cite{scheldrick2015shelxt}.

\section*{Acknowledgements}
We thank Nobuyoshi Miyajima for help with the FIB milling. 
\section*{Funding}
We thank the German Research Foundation (Deutsche Forschungsgemeinschaft, DFG, Project Nos. DU 954/11-1, DU 393/13-1, DU 393/9-2, STE 1105/13-1 and ME 5206/3-1) and the Federal Ministry of Education and Research, Germany (BMBF, Grant No. 05K19WC1) for financial support. TM was supported by the Center for High Pressure Science and Technology Advance Research, Beijing, P.R. China.
FT was supported by the Swedish Research Council (VR) Grant No. 2019-05600. D.L. thanks the Alexander von Humboldt Foundation for financial support. N.D. thanks the Swedish Government Strategic Research Area in Materials Science on Functional Materials at Linköping University (Faculty Grant SFO-Mat-LiU No. 2009 00971).
\\

%

\end{document}